# Large area and structured epitaxial graphene produced by confinement controlled sublimation of silicon carbide

Walt. A. de Heer[1], Claire Berger[1,2], Ming Ruan[1], Mike Sprinkle[1], Xuebin Li[1], Yike Hu[1], Baiqian Zhang[1], John Hankinson[1], Edward H. Conrad[1]

[1] Georgia Institute of Technology, School of Physics, Atlanta, GA 30332-0430, USA
[2] CNRS- Institut Néel, Grenoble, France

ABSTRACT

After the pioneering investigations into graphene-based electronics at Georgia Tech (GT), great strides have been made developing epitaxial graphene on silicon carbide (EG) as a new electronic material. EG has not only demonstrated its potential for large scale applications, it also has become an invaluable material for fundamental two-dimensional electron gas physics showing that only EG is on route to define future graphene science. It was long known that graphene mono and multilayers grow on SiC crystals at high temperatures in ultra-high vacuum. At these temperatures, silicon sublimes from the surface and the carbon rich surface layer transforms to graphene. However the quality of the graphene produced in ultrahigh vacuum is poor due to the high sublimation rates at relatively low temperatures. The GT team developed growth methods involving encapsulating the SiC crystals in graphite enclosures, thereby sequestering the evaporated silicon and bringing growth process closer to equilibrium. In this confinement controlled sublimation (CCS) process, very high quality graphene is grown on both polar faces of the SiC crystals. Since 2003, over 50 publications used CCS grown graphene, where it is known as the "furnace grown" graphene. Graphene multilayers grown on the carbon-terminated face of SiC, using the CCS method, were shown to consist of decoupled high mobility graphene layers. The CCS method is now applied on structured silicon carbide surfaces to produce high mobility nano-patterned graphene structures thereby demonstrating that EG is a viable contender for next-generation electronics. Here we present the CCS method and demonstrate several of epitaxial graphene's outstanding properties and applications.

**Introduction**

Graphene has been known and studied for decades in many forms, [1,2,3] but it was not until 2001 when graphene's potential for electronics was recognized [4,5]. Graphene-based electronics requires a patternable form of graphene with excellent electronic properties. It was clear from the outset [4] that epitaxial graphene on silicon carbide (EG) [2] was the most promising form of graphene for this project. However, EG produced by conventional methods is of poor quality[6]. It was clear that much higher quality material would need to be produced for graphene-based electronics to have any chance to be realized. The confinement controlled sublimation (CCS) method that is described here is well on its way to satisfy multiple stringent conditions required for a viable electronic material.

Note that the somewhat later parallel development of back-gated exfoliated graphitic flakes deposited on oxidized silicon[7], has little practical impact in the development of graphene electronics. While back-gated graphene flakes are extremely useful for two dimensional electron gas research, randomly deposited graphitic flakes obviously do not constitute an electronic material and back-gating (that switches all devices on the substrate at once) is not useful for electronics.

**Production of epitaxial graphene**

Van Bommel first showed in 1975 that a graphene layer grows on hexagonal silicon carbide in ultra high vacuum (UHV) at temperatures above about 800 C.[2] Silicon sublimation from the SiC causes a carbon rich surface that nucleates an epitaxial graphene layer, Fig. 1. The graphene growth rate was found to depend on the specific polar SiC crystal face: graphene forms much slower on the silicon-terminated face (0001) surface (or Si-face) than on the carbon-terminated face (000-1) surface (or C-face). Van Bommel identified *monocrystalline graphite monolayer films* (i.e. graphene)[2] that were found to be essentially decoupled from the SiC substrate[6] and therefore were electronically equivalent to isolated graphene sheets.[1] Since 1975, these films were referred to as monolayer graphite, or two-dimensional graphite crystals or epitaxial graphene. In fact, electronically decoupled epitaxial graphene had been observed on many surfaces, such as Pt, Ni, Ru, Ir etc[3], but until 2001 the electronic properties and the applications potential of graphene had not been considered. By 2003 the ideas were fully developed and backed with compelling scientific evidence (that was published[5] in 2004). The invention of graphene based electronics was patented in June 2003.[4]

The first graphene transport measurements were performed on epitaxial graphene films grown by the UHV sublimation method[5, 8]. Graphene films produced this way are defective[6] (Fig. 2A) and have low mobilities[5] ($\mu \sim 15$ $cm^2V^{-1}s^{-1}$). Nonetheless, investigations demonstrated that these epitaxial graphene films could be patterned using standard microelectronics methods and that the films had two-dimensional electron gas properties. By 2003 the UHV sublimation process had been improved to produce monolayer graphene films with mobilities exceeding $10^3$ $cm^2V^{-1}s^{-1}$.[5] Magnetotransport measurements of these films showed precursors of the half-integer quantum Hall effect with its characteristic non-trivial Berry's phase[9]. While the improvement was significant, these mobilities are still low compared with nanotubes[10] or graphite[11].

Defects in UHV sublimed silicon carbide can be traced to the relatively low growth temperatures and the high graphitization rates in the out-of equilibrium UHV sublimation process. While increased growth temperature will anneal vacancies and grain boundaries, the UHV growth method still leads to unacceptable high sublimation rates. There are a number of way to control the rate at which silicon sublimes. For example by supplying silicon in a vapor phase compound (e.g., silane[12]) or by flowing an inert gas over the hot silicon carbide surface[13].

Alternatively, the confinement controlled sublimation method developed at GT relies on confining the silicon carbide in a graphite enclosure (either in vacuum or in an inert gas). This limits the escape of Si and thus maintains a high Si vapor pressure so that graphene growth proceeds close to thermodynamic equilibrium (Fig. 1B). Graphene growth over macroscopic areas can be controlled on both polar faces of SiC to produce either monolayer graphene or multilayer graphene films (Fig. 1E, F). Over 2000 individual CCS epitaxial graphene samples have been made and CCS produced graphene has been characterized in over 50 publications, where it is referred to as *furnace grown epitaxial graphene*. Particularly relevant examples are the demonstration of infra-red Landau level spectroscopy showing very high moblities [14], quantum Hall effect [15], scanning tunneling Landau level spectroscopy [16], fractional Landau level filling factors [17], and self assembly of graphene ribbons [18] large scale patterning, electronic confinement and coherence [19], electronic structure of decoupled layers in multilayered epitaxial graphene [20, 21]

Until now, details of the CCS method have not been published. The principle of CCS can be understood from kinetic gas theory [Fig. 1A, B]. Graphene growth is proportional to the rate of silicon depletion from the SiC surface, since each evaporated silicon atom leaves behind one carbon atom on the surface. In thermodynamic equilibrium the Si evaporation rate, $n^-$, and the Si condensation rate, $n^+$ at the SiC are exactly balanced so that graphene does not form. This condition will eventually be established in a hermetically sealed, non-reactive, enclosure at any temperature, after the enclosure surfaces have been passivated. In our design, we use a graphite enclosure and passivation of the enclosure is achieved after several graphene growth cycles. In more detail, assume that a Si atom impinging on the surface condenses with a sticking probability ε, ( $0 \leq \varepsilon \leq 1$ ) then $n^+ = \varepsilon\, v_{ave}\, \rho_{eq}/4$ where $v_{ave} = \sqrt{(8kT/\pi m)}$ is the average thermal speed of a silicon atom in the vapor, $m$ is its atomic mass and $\rho_{eq}(T)$ is the vapor density of silicon in equilibrium with silicon carbide at temperature $T$. Consequently $n^+ \approx \varepsilon\, \rho_{eq}(2kT/\pi m)^{1/2}$. The sticking coefficient (but not the vapor density) depends on the local surface structure, and the polar face. For simplicity, in the rest of this discussion we assume $\varepsilon=1$, independent of T. However we note that graphene growth rates are greater on the C-face than on the Si face, implying that $\varepsilon$ is greater on the former than on the latter which is important for certain implementations of the method. Clearly, the silicon must escape through the layers that have already formed, so that the rates depend on the graphene thickness. It appears however, that for thinner layers, silicon manages to readily escape from the silicon carbide surface.

If the enclosure is not hermetically sealed, but supplied with a small calibrated leak [Fig. 1B], then $n^- > n^+$ causing graphene to grow at a rate $n^{gr} = n^- - n^+$. Consequently, $n^{gr}$ is controlled by the size of the leak. In general, the rate at which silicon atoms escape is $N = C\, v_{ave}\, \rho_{eq}$, where $C$ is the effective area of the leak (for a cylindrical hole of diameter $D$ and length $L$, $C=D^3/3L$). Consequently, $n^{gr} = N/A$ where A is the crystal surface area. For example, for a 1 cm$^2$ crystal in vacuum, with $L = 1$ cm and $D = 0.75$ mm the graphene formation rate is reduced by more that a factor of about 1000 compared to the UHV sublimation method (in which $n^+ = 0$).

The actual rates can be estimated from the vapor pressure of $P_{Si}(T)$ of Si over SiC, as has been determined by Lilov [22]: $P_{Si}$ (1500K)=1.7 $10^{-6}$ Torr, $P_{Si}$ (2000K)=1.1 $10^{-2}$ Torr, $P_{Si}$ (2500K)=1.4 Torr, consequently, $P_{Si}$ and $\rho_{eq}$ increase by about a factor of 7 per 100K. Assuming the sticking coefficient for the C-face is $\varepsilon=1$, and that one carbon atom remains for every evaporated silicon atom, then a graphene monolayer forms on the C-face in about one minute at T=1200 C for a SiC crystal that freely sublimes in vacuum. This formation rate reasonably agrees with the experimental graphene formation rate in UHV. Consequently, compared with the UHV sublimation method, the CCS method allows the sample temperature to be increased by about 300 K for a given rate of graphene growth. This has been experimentally confirmed for the enclosure described above.

Introducing an inert gas further decreases the growth rate. In that case, silicon atoms must diffuse through the gas filled leak in order to escape the enclosure. This reduces the Si leak rate by a factor $R = (D/\lambda +1)^{-1}$ where $\lambda$ is the mean free path of a silicon atom in the gas (see, for example Ref. 23). For example, for argon, $\lambda=(\sigma_{Ar-Si}\rho_{Ar})^{-1}$ where $\rho_{Ar}$ is the Ar density and $\sigma_{Ar-Si}\approx$ 30 $Å^2$ is the estimated [23] Ar-Si gas kinetic scattering cross section so that for $P = 1$ Bar, $R \approx 10^{-3}$ in the example above. Hence, the graphene formation rates can be reduced by an additional factor of up to $10^3$ by introducing argon into the enclosed volume. Consequently, the CCS method allows growth rates to be adjusted over a factor of $10^6$ compared with UHV growth. Moreover, the growth temperature and the growth rates can be independently tuned: coarsely tuned by the leak out of the confinement volume and finely tuned by introducing an inert gas. Compare to the *Edison Lightbulb method* introduced by Emtsev et al. [13], which only uses flowing argon to restrict Si sublimation, the CCS method is more flexible. Furthermore, the *Lightbulb* method is intrinsically far from equilibrium and its effectiveness for C-face graphene growth has not been demonstrated.

**The two varieties of EG**

Van Bommel first observed the differences between graphene grown on the silicon (0001) and the carbon (000-1) terminated faces of hexagonal silicon carbide. [2] Low energy electron diffraction (LEED) and angle resolved photoemission spectroscopy (ARPES) reveals that Si-face graphene monolayers exhibit the characteristic linear bandstructure (a.k.a. Dirac cones). Typical monolayer mobilities using CCS on the Si-face are found to be modest and typically on the order of $10^3$ $cm^2V^{-1}s^{-1}$. Like in graphite, Si-face graphene multilayers are Bernal stacked; Si-face grown bi-layers exhibit parabolic bands and with increasing thickness, the band structure evolves to that of graphite [24]. Consequently Si-face graphene multilayers are actually ultrathin graphite films and known as *few layer graphite* or FLG (this acronym is also inaccurately interpreted as few layer graphene).

Van Bommel also observed that UHV grown graphene on the C-face is both rotationally disordered and defective. [2] However, C-face graphene grown by the CCS method shows rotational order, consisting primary of two principle rotational orientations (Fig. 2F), in contrast to the single orientation in Bernal stacked graphene and Si-face FLG (Fig. 2D). While the exact structure is not known, it is consistent with a stacking where every other layer is aligned within ±~7 degrees of the SiC <21-30> direction and separated by layers rotated by 30° with respect to the <21-30> direction (Bernal stacked layers make up no more than 15% of the film and are considered stacking faults in this structure). [25] An important consequence of this stacking is that each graphene layer in the stack has the same electronic structure as an isolated graphene sheet and therefore behaves as if it is electronically decoupled from its neighbor. Therefore, C-face multilayers produced by the CCS method are multilayer epitaxial graphene (MEG) [8, 21] and not thin graphite (or FLG). This important property has been confirmed by a variety of probes. For instance the Raman spectrum of the ~100 layer MEG sample of Fig. 3 shows the characteristic G and single Lorentzian 2D peaks of single layer graphene. More specifically, ARPES was used to directly image the linear graphene band structure [20] [see Fig. 2]. In addition, optical transitions between Landau levels in MEG have been observed even at room temperature in low magnetic fields, indicating very weak electron phonon coupling and room temperature mobilities exceed 250,000 $cm^2V^{-1}s^{-1}$ for interior MEG layers [14]. These features are clearly important for graphene science. Recent low temperature high magnetic field scanning probe investigations have directly imaged the quantum Hall states in MEG [16]. This work also demonstrated that MEG layers are atomically flat (with 50 pm height variations) between substrate steps (that can be up to 50 μm apart).

An important property of both varieties of EG produced by the CCS method, is that the graphene layers are continuous over substrate steps; the morphology is likened to a carpet that is draped over the SiC surface [8]. In fact scanning tunneling microscopy has not revealed a break in the top graphene layer. Hence, at least the top MEG layer covers the entire surface of a macroscopic SiC wafer.

The graphene/SiC interface on the Si-face is well understood and is defined by a non-conducting carbon rich buffer layer with a 6√3x6√3 structure that causes a corrugation between 0.5-0.8Å of the first graphene layer.[26, 1] The C-face interface is less well understood. Surface x-ray diffraction indicates that it is also carbon rich with a density close to diamond, while LEED from the thinnest C-face films (discussed below) reveals only a (1x1) pattern. The interface is found to be well defined and flat, consistent with a carbon rich layer at the interface that is tightly bound to the underlying SiC. [26]

On both faces, the graphene/SiC interface is charged, inducing a negative charge density $n^{gr} \approx 5\times10^{12}\text{-}10^{13}$ $cm^{-2}$ on the first graphene layer. ARPES and transport measurements show that this layer (C-face) or the layer just above it (Si-face) in CCS produced epitaxial graphene has the characteristic linear graphene dispersion and high mobility (graphene ribbon mobilities are 500-2000 $cm^2$ $V^{-1}$ $s^{-1}$ for the Si-face and

10,000-30,000 cm$^2$ V$^{-1}$ s$^{-1}$ for the C-face) [8]. The interior layers in MEG are essentially neutral (a screening length of about 1 layer has been measured [27, 28]).

Bernal stacking of graphene layers on the Si-face may explain its low mobilities (compared with the C-face). This may be caused by a graphite bilayer gap that develops from an embryonic second layers that grows at step edges under a completed top graphene layer [13]. Since the local bi-layer electronic structure is a significant perturbation to the graphene electronic structure, scattering there will be important. In contrast, because of the electronic decoupling, this scattering mechanism is likely to be suppressed in C-face graphene. The scattering from partially formed layers at the interface is therefore less important for C-face graphene compared with Si-face, explaining the observed large mobility differences. Although electronic scattering mechanisms in graphene are not well understood, it is expected that interface disorder is a significant factor in epitaxial graphene. Interface passivation and annealing processes are being developed to further increase mobilities [29].

**Applications of the CCS method**

**Large area graphene growth**

The CCS method is routinely used to cover the entire surface of an on-axis or off-axis cut silicon carbide chips (4H or 6H, Si-face or C-face) with a graphene monolayer or multilayer. These chips are used to measure EG properties or as a starting material for graphene device structures.

The various stages of the graphitzation process of a 20 μm x 20 μm region of a 6H *Si-face* chip are demonstrated in Fig. 4. The hydrogen-etched surface (Fig. 3A) exhibits characteristic half- unit cell silicon carbide steps (0.8 nm) that result from the miscut. The chip was subsequently heated to several temperatures (Fig. 4B-D). At 1300 C, the steps become rounded and at 1400 C, they roughen. The roughing is accompanied by the formation of the buffer layer as verified in LEED and observed to occur at T=1080 C in UHV grown epitaxial graphene. [30] Note that the graphitization temperature increases by ~300 C in the CCS method compared to the UHV method is consistent with the kinetic model of the CCS process. At 1520 C, a high-quality graphene monolayer forms on the C-face in about 20 min (Fig. 5A). In contrast, a defective monolayer forms in a few minutes at 1250 C in UHV [31]. It should be noted that AFM images gives the illusion of significant substrate roughness. However, the typical step height is about 1 nm that is about 1/1000 of a typical terrace width. Moreover, the graphene mobilities are high (typically >10,000 cm$^{2/}$Vs on the C-face), even for monolayers [15]. Hence, the substrate steps (at least on the C-face) appear not to be a significant source of scattering.

Figure 5 shows a 1 cm x 1cm C-face 4H chip that was CCS graphitized. The C-face was graphitized for 10 min at 1550 C to produce an essentially uniform graphene monolayer over the entire surface as verified in ellipsometry measurements (Fig. 5C) (ellipsometer spot size 250μm x 250μm). LEED shows a single set of diffraction spots, consistent with a single graphene layer, oriented 30° with respect to the SiC lattice (cf Fig 5D). Raman spectroscopy shows characteristics of defect free thin graphene: a narrow 2D peak single Lorentzian centered at 2701cm$^{-1}$, (FWHM = 28 cm$^{-1}$) and no significant D peak indicating high quality graphene (Fig. 5B). The 2D and G peaks are blue shifted compared with exfoliated graphene flakes on SiO$_2$ (by ~6cm$^{-1}$ and 20cm$^{-1}$ respectively), but the shift is smaller than for epitaxial graphene on the Si-face. The peaks position does not vary significantly with the number of layers, and a single Lorentzian 2D peak is primarily observed on very thick MEG films as a result of the electronic decoupling. The blue shift observed in Si face graphene is attributed to strain due to a differential thermal contraction upon cooling between the SiC substrate and the graphene [32]. The small shift observed here indicates a reduced strain in C-face grown mono and multilayers. Sample to sample variability is observed in the magnitude of the shifts and peak intensities. The expected attenuation of the SiC signal correlates well with sample thickness, allowing a crude sample thickness measurement from Raman data with a precision of about 2-3 layers.

Atomic force microscopy (AFM) images like Fig. 5A show the typical pleat structure of the C-face epitaxial graphene layer (pleats are not typically seen on Si-face graphene). The pleats (also called puckers, ridges, creases, rumples, ripples and folds) are typically 1-10 nm high, are typically spaced 3-10 µm apart and are thought to result from the differential expansion of the silicon carbide and graphene and the very weak coupling of the graphene to the substrate. Room temperature van der Pauw transport measurements over the 1 cm x 1cm area give a Hall mobility of 2000 cm$^2$ V$^{-1}$ s$^{-1}$. We find that transport measurements over cm size scales typically give significantly lower values than measurements of micron sized Hall bars, which may be related to residual inhomogeneities over macroscopic length scales. In contrast, STM studies show that graphene structure is continuous over the pleats and transport measurement show that pleats do not significantly affect the transport, in fact high mobility monolayer graphene Hall bars ($\mu$~20,000 cm$^2$ V$^{-1}$ s$^{-1}$) with many pleats show a well-developed quantum Hall effect and no indication of non-uniform transport [15].

Large area graphene multilayers are grown both on the C-face and the Si-face by increasing the annealing times and the temperatures. For large area C-face growth, temperatures in the range of 1450-1525 C are preferred because at these reduced temperatures, step bunching is reduced so that large substrate steps are avoided. Note however, that over macroscopic distances, the unavoidable miscut of a nominally flat SiC crystal (of order 0.05°) always results in substrate steps.

**Growing graphene on mesas**

One way to control and to eliminate substrate steps in defined regions of the silicon carbide substrate, is to etch mesas on the surface. The subsequent heat treatment will cause the substrate steps to flow and bunch at the mesa boundaries and ultimately produce step-free SiC surfaces.

This process relates to earlier research that demonstrated step-free mesas in heteroepitaxial SiC step-flow growth on vicinal SiC, where step pinning occurs at the step edges [33]. The near equilibrium growth with correspondingly higher kinetics in the CCS process goes beyond step flow growth to approach near equilibrium crystal shapes with essentially atomically flat surfaces. Subsequent graphitization produces atomically flat graphene (except for the pleats). Growth on mesas is demonstrated on hexagonal and square mesas, 2 micron high, that were lithographically etched on the C-face of a 4H-SiC chip. The samples were heated to 1250 C for 20 min and subsequently to 1600 C for 10 min. Mesas smaller than 20 μm generally became flat and the edges slope at a 57 deg angle from the vertical. While substrate steps are still observed on these surfaces, their densities are at least a factor of 10 smaller than the on the original hydrogen etched SiC surface. We have further found that graphene formation inhibits step bunching. Therefore the mesas were produced by annealing the structures twice and graphene layer was etched away (using an oxygen plasma etch) between the annealing steps. Figure 6 B shows part of a 100 μm x 100 μm mesa, that corresponds to essentially a single SiC terrace (cf. Fig. 5 A). Conditions are currently sought under which the graphene growth is so slow that the SiC surfaces anneal to their equilibrium shapes before they graphitize. In that case mesa flattening and graphene growth can be accomplished in one step.

**Structured graphene growth**

Extended graphene sheets are important for basic research. However, essentially all electronic applications require graphene to be patterned. In fact, graphene electronics relies on the possibly to interconnect nanoscopic graphene structures. An extended graphene sheet is a gapless semimetal but a graphene ribbon is expected to develop a bandgap $\Delta E \approx 1 eV/W$, where W is the width of the ribbon in nm [34, 35]. Graphene ribbons can be patterned using standard microelectronics lithography methods [36]. In these methods, an extended graphene sheet is coated with an electron-beam resist material (for example, PMMA) that is patterned by e-beam lithography. After development of the resist, the graphene is oxygen ion-etched to produce the desired structures. It is clear that the lithography steps themselves are destructive and produce non-ideal edges and poorly defined structures at the nanoscale. This contributes to significantly reduced mobilities and spurious electronic localization effects [35] In fact the mobility reduction in conventionally patterned structures is so severe that the nanoelectronics applications potential of graphene has been called into question. It is clearly advantageous to avoid processing steps that involve cutting the graphene. A nuanced approach that avoids nanolithography of the graphene is clearly desired.

A very promising method is to grow graphene on structured silicon carbide. In this method, the ungraphitized Si-face (0001) of silicon carbide is lithographically patterned in the usual way using a resist coating. The exposed SiC areas are then plasma etched using $SF_6$ or $CF_4$ so to produce depressions of well-defined depths ranging from a few nm to microns as controlled by the intensity and duration of the plasma etching procedure. The sample is finally annealed and graphitized by the CCS method at temperatures typically in the range of 1550 C to 1650 C.

In this process, the sidewalls of the etched structures crystallize, typically along the (1-10n) direction, n=8 for 4H SiC and n=12 for 6H SiC. Consequently a circular etched mesa with a diameter of 1 µm will crystallize into hexagons (Fig. 6c) where the sidewall slopes are about 62 deg from the vertical. The graphitization rates of the (1-10n) sidewalls are similar to the graphitization rates of the (000-1) surfaces (i.e. the C-face) of SiC. Because these rates are much greater than the graphitization rates of the Si-face horizontal surfaces, only the sidewalls and not the (0001) flat surfaces are graphitized. Consequently, this graphitization method can be controlled to produce monolayer graphene on the sidewalls while graphene on the horizontal (0001) surfaces are sub-monolayer and non-conducting. This has been verified by Raman spectroscopy and by transport measurements.

Along the same lines, graphitization on natural (non pre-etched) vicinal steps of SiC on-axis or off-axis (4 degree or 8 degree from (0001)) produce arrays of narrow ribbons. Significant step bunching precedes the graphitization to produce step heights on the order of 10 nm with corresponding ribbon widths.

The process has been applied to more complex structures as shown in Fig. 7A-C, where a two vertical 400 nm deep trenches are crossed with a 15 nm deep, 1µm wide depression. The structure is annealed and graphitized using the CCS method. Annealing causes the observed rounding of the previously straight sidewalls and graphitization produces a 40 nm graphene sidewall ribbon as revealed by electrostatic force microscopy. The increased width compared with the sidewall width is caused by recrystallization of the vertical sidewalls (initially exposing a 1-100 face) to (1-10n) with a reduced slope (23 deg from horizontal for n =10) to produce wider ribbons.

The narrow ribbons are seamlessly connected to wide ribbons that were formed on the sidewalls of the deep trench as revealed in the electrostatic force micrograph (Fig. 7C). Transport measurements demonstrate the continuity of the sidewall ribbons. The resistivity of sidewall ribbons is about 100 _ per square, which is comparable to measurements on doped two dimensional epitaxial graphene sheets (n doping on the order of $10^{12}/cm^2$ is typically observed). These resitivities are orders of magnitude smaller than observed in lithographically patterned ribbons of comparable width, indicating that the sidewall graphitization method produces ribbons with smoother edges. Moreover, there is no evidence for Coulomb blockade effects that result from edge roughness in exfoliated graphene ribbons [37]. In fact, recent measurements of gated sidewall ribbons between 20 and 40 nm wide and from 1 to 2 µm long indicate that they are ballistic, resembling carbon nanotubes in that respect. In fact, there is no indication of a bandgap in these ribbons (these results will be published elsewhere). These examples show that sidewall graphitization is an effective and simple method to produce interconnected nanoscopic graphene ribbons and other graphene nanostructures, without resorting to nanolithographic methods.

**Summary and conclusion**

The near equilibrium, confinement controlled sublimation method to produce epitaxial graphene (mono and multi) layers on silicon carbide has been demonstrated to be a versatile method to produce high quality uniform graphene layers on both the Si-face and the C-face of single crystal silicon carbide. It provides control of the silicon vapor density and assures that the density is constant over the surface and near thermodynamic equilibrium, which is essential for uniform growth. The method allows good control of the graphitization temperatures, which is important, since growth at low temperatures (as in the case of sublimation in unconfined ultra-high vacuum) produces defective graphene layers. The CCS method allows further control of the graphitization rates by introducing inert gasses, which can essentially inhibit the graphene growth even at temperatures exceeding 1600 C. This is important if the graphitization is preceded by an annealing step of the silicon carbide surface itself for example to anneal (or recrystallize) a structured silicon carbide surface.

Graphene electronics imposes great demands on the material and material processing. It should be clear that the realization of graphene-based electronics requires all of these conditions to be met. The CCS method is an important step in the production of high quality graphene both in single layers and multilayers. Sidewall graphitization by the CCS method has recently been advanced to the point that narrow ballistic graphene ribbons can be made. This is an important development that overcomes the edge roughness, which is the most serious lithography problem. Hence, graphene ribbons are approaching the electronic quality of carbon nanotubes. Furthermore, beyond graphene growth, great advances have been made in producing effective top gate structures that do not introduce additional scattering. Despite all these advances, graphene based electronics has not yet been realized. However given progress in epitaxial graphene on silicon carbide, the prospects are encouraging.

**Acknowledgements**

The authors thank the W. M Keck foundation, the National Science Foundation (DMR-0820382), the Air Force Office of Scientific Research for financial support and the Partner University Fund for travel grant.

**References**

1. Boehm, H., Clauss, A., Hofmann, U & Fischer, G. Dunnste Kohlenstof-Folien. *Z Fur Naturforschung* **B 17**, 150 (1962).
2. Van Bommel, A., Crombeen, J. & van Tooren, A. LEED and Auger-electron observations of SiC (0001) surface. *Surface Science* **48**, 463-472 (1975).
3. Gall, N.R., RutKov, E.V. & Tontegode, A.Y. Two dimensional graphite films on metals and their intercalation. *Int. J. of Mod. Phys. B* **11**, 1865-1911 (1997).

4. de Heer W.A., Berger, C., First, P.N. United States Patent US 7015142, filed June 2003, issued 2006; de Heer, W.A. Early Development of Graphene Electronics. (2009) at <http://hdl.handle.net/1853/31270>
5. Berger, C. et al. Ultrathin epitaxial graphite: 2D electron gas properties and a route toward graphene-based nanoelectronics. *J. Phys. Chem. B* **108**, 19912-19916 (2004).
6. Forbeaux, I., Themlin, J.M. & Debever, J.M. Heteroepitaxial graphite on 6H-SiC(0001): Interface formation through conduction-band electronic structure. *Phys. Rev. B* **58**, 16396-16406 (1998).
7. Novoselov, K.S. et al. Electric Field Effect in Atomically Thin Carbon Films. *Science* **306**, 666-669 (2004).
8. de Heer, W. et al. Epitaxial graphene. *Solid State. Com.* **143**, 92-100 (2007).
9. Ando, T., Fowler, A.B. & Stern, F. Electronic properties of two-dimensional systems. *Reviews of Modern Physics* **54**, 437-672 (1982).
10. McEuen, P.L., Fuhrer, M.S. & Park, H.K. Single-walled carbon nanotube electronics. *IEEE Transactions on Nanotechnology* **1**, 78-85 (2002).
11. Soule, D. & McClure, J. Band Structure And Transport Properties Of Single-Crystal Graphite. *J. Phys. Chem. of Solids* **8**, 29-35 (1959).
12. Tromp, R. & Hannon, J. Thermodynamics and kinetics of graphene growth on SiC(0001). *Phys. Rev. Lett.* **102**, 106104 (2009).
13. Emtsev, K. et al. Towards wafer-size graphene layers by atmospheric pressure graphitization of silicon carbide. *Nature Mat.* **8**, 203-207 (2009).
14. Orlita, M. et al. Approaching the Dirac Point in High-Mobility Multilayer Epitaxial Graphene. *Phys. Rev. Lett* **101**, 267601-4 (2008).
15. Wu, X. et al. Half integer quantum Hall effect in high mobility single layer epitaxial graphene. *App. Phys. Lett.* **95**, 223108 (2009).
16. Miller, D.L. et al. Observing the Quantization of Zero Mass Carriers in Graphene. *Science* **324**, 924-927 (2009).
17. Song, Y. et al. High-resolution tunnelling spectroscopy of a graphene quartet. *Nature* **467**, 185-189 (2010).
18. Sprinkle, M. et al. Scalable templated growth of graphene nanoribbons on SiC. *Nature Nanotech* **5**, 727-731 (2010).
19. Berger, C. et al. Electronic confinement and coherence in patterned epitaxial graphene. *Science* **312,** 1191 (2006)
20. Sprinkle, M. et al. First direct observation of a nearly ideal graphene band structure. *Phys. Rev. Lett.* **103**, 4 (2009).
21. Hass, J. et al. Why multilayer graphene on 4H-SiC(000$\bar{1}$) behaves like a single sheet of graphene. *Phys. Rev. Lett.* **100**, 125504 (2008).
22. Lilov, S. Study of the Equilibrium Processes In The Gas-Phase During Silicon-Carbide Sublimation. *Mat. Science and Engineering B-* **21**, 65-69 (1993).
23. Hoffman, D., Singh, B. & Thomas, I.J.H. *Handbook of Vacuum Science and Technology*. (Academic Press: 1998).
24. Zhou, S.Y. et al. Substrate-induced bandgap opening in epitaxial graphene. *Nature Materials* **6**, 770-775 (2007).

25. Sprinkle, M. et al. Multilayer epitaxial graphene grown on the SiC (000-1) surface; structure and electronic properties. *J. Phys. D* **43**, 374006 (2010).
26. Hass, J., Millan-Otoya, J., First, P. & Conrad, E. Interface structure of epitaxial graphene grown on 4H-SiC(0001). *Phys. Rev. B* **78**, 205424 (2008).
27. Sun, D. et al. Spectroscopic Measurement of Interlayer Screening in Multilayer Epitaxial Graphene. *Phys. Rev. Lett.* **104**, 136802 (2009).
28. Sun, D. et al. Ultrafast Relaxation of Excited Dirac Fermions in Epitaxial Graphene Using Optical Differential Transmission Spectroscopy. *Phys. Rev. Lett.* **101**, 157402-4 (2008).
29. Riedl, C., Coletti, C., Iwasaki, T., Zakharov, A.A. & Starke, U. Quasi-free Standing Epitaxial Graphene on SiC by Hydrogen Intercalation. *Phys. Rev. Lett.* **103**, 246804 (2009).
30. Emtsev, K., Speck, F., Seyller, T., Ley, L. & Riley, J. Interaction, growth, and ordering of epitaxial graphene on SiC{0001} surfaces: A comparative photoelectron spectroscopy study. *Phys. Rev. B* **77**, 155303 (2008).
31. Mallet, P. et al. Electron states of mono- and bilayer graphene on SiC probed by scanning-tunneling microscopy. *Phys. Rev. B* **76**, 041403 (2007).
32. Rohrl, J. et al. Raman spectra of epitaxial graphene on SiC(0001). *Appl. Phys. Lett.* **92**, 201918-3 (2008).
33. Powell, J. et al. Growth of step-free surfaces on device-size (0001)SiC mesas. *Appl. Phys. Lett.* **77**, 1449-1451 (2000).
34. Nakada, K., Fujita, M., Dresselhaus, G. & Dresselhaus, M.S. Edge state in graphene ribbons: Nanometer size effect and edge shape dependence. *Phys. Rev. B* **54**, 17954-17961 (1996).
35. Han, M.Y., Ozyilmaz, B., Zhang, Y. & Kim, P. Energy Band-Gap Engineering of Graphene Nanoribbons. *Phys. Rev. Lett.* **98**, 206805 (2007).
36. Li, X. et al. Top- and side-gated epitaxial graphene field effect transistors. *Phys. Stat. Solid.* **207**, 286-290 (2010).
37. Ozyilmaz, B., Jarillo-Herrero, P., Efetov, D. & Kim, P. Electronic transport in locally gated graphene nanoconstrictions. *Appl. Phys. Lett.* **91**, 192107 (2007).

**Figure Captions**

Figure 1:

The Confinement Controlled Sublimation method. (A) SiC wafer in UHV: sublimed silicon is not confined, causing rapid, out of equilibrium graphene growth. (B) The CCS method: sublimed Si gas is confined in a graphite enclosure so that growth occurs in near thermodynamic equilibrium. Growth rate is controlled by the enclosure aperture (leak), and the background gas pressure. (C) Photograph of the induction furnace. (D) under CCS conditions few layer graphite (FLG, from 1 to 10 layers) grows on the Si terminated face, and multilayer epitaxial graphene (MEG, from 1 to 100 layers) grows on the C terminated face.

Figure 2:

Comparison of UHV and CCS grown epitaxial graphene. (A-C) AFM images and (D-F) LEED patterns. (A) UHV grown monolayer on the Si-face. (B) CCS monolayer grown on Si-face. (C) MEG on C face; note that layers drape over the substrate steps (white lines are pleats in the MEG layer). (D) LEED pattern of CCS grown Si-face graphene monolayer (bright spots due to graphene) showing typical surface reconstruction features. (E) LEED image of CCS grown C-face monolayer; (F) LEED image of CCS grown C-face multilayer (MEG), showing characteristic "arcs" due to the rotational stacking.

Figure 3:

Raman and ARPES spectroscopy of MEG (left and right panels, resp.). Raman spectrum shows the characteristic G and 2D graphene peaks. The 2D peak of this ~100 layer MEG sample fits a single Lorentzian of full width 25cm$^{-1}$, centered at 2701.8 cm$^{-1}$. A weak D peak can be discerned. ARPES data from the top three layers of a 10 layer sample around $k_y$=0 (i.e., the Dirac point). Two unperturbed cones are observed showing that the layers are electronically decoupled: the slope is $v_F$=1x10$^6$ m/s, as expected for isolated graphene, and the Fermi level is with 20 meV from the Dirac point. For details, see Ref. 20.

Figure 4:

AFM images showing the evolution of the surface of the 6H Si-face upon annealing. (A) initial surface after hydrogen etching showing half-unit cell steps (0.8 nm) resulting from the miscut; (B) After CCS annealing at 1300 C: substrate steps become rounded, (C) annealing at 1400C; the steps roughen, and (D) 1500 C: formation of a graphene layer. The scale bar is 5 μm.

Figure 5:

Single layer graphene grown on a 1cm x 1cm 4H-SiC chip on the C-face. (A) AFM image showing a graphene coated stepped SiC surface; note the continuous pleats running across large regions of the surface. (B) The Raman spectrum (after SiC background subtraction); the 2D peak consists of single Lorentzian (full width =28 cm$^{-1}$); the disorder induced D peak is absent. (C) Ellipsometry shows graphene uniformity (color scale light blue: 1 layer, yellow: 2 layers, red: 3 layers, dark blue: no graphene; beam size : 250μm x 250μm). (D) LEED pattern of the monolayer.

Figure 6:

Examples of template grown graphene structures etched on the (0001) face. (A) 4H-SiC hydrogen-etched surface with a regular step structure. (B) Flat step-free graphitized mesa with MEG pleats (C) circular mesa etched on Si-face; the hexagonal shape results from the annealing at 1550 C showing preference for the (1-10n) crystal surfaces (n depends on the step height and ranges from 2, for nm steps to about 10 for μm steps). (D) Electrostatic force microscopy image after CCS annealing; graphene (light) grows on the mesa sidewalls but not on the horizontal (0001) surfaces.

Figure 7:

Templated CCS sidewall graphene ribbons. (a) AFM image of two 1μm deep trenches (brown) etched in SiC Si-face (scale bars are 1μm). The trenches are connected by a 10 nm deep trench (light brown). (b) After annealing, the initially straight trench walls becomes rounded, indicating significant SiC mass flow. (c) EFM shows that 20 nm graphene graphene ribbons have grown on the trench sidewalls (verified in Raman spectroscopy). The wide graphene ribbons concurrently grown on deep μm high sidewalls serve as current leads and voltage probes. (d) I-V characteristics of the two 20 nm wide, 1.5 μm long ribbons in parallel. The resistance corresponds to 8 kΩ per ribbon (corresponding to about 100 Ω/square) and exhibits a weak zero bias anomaly.

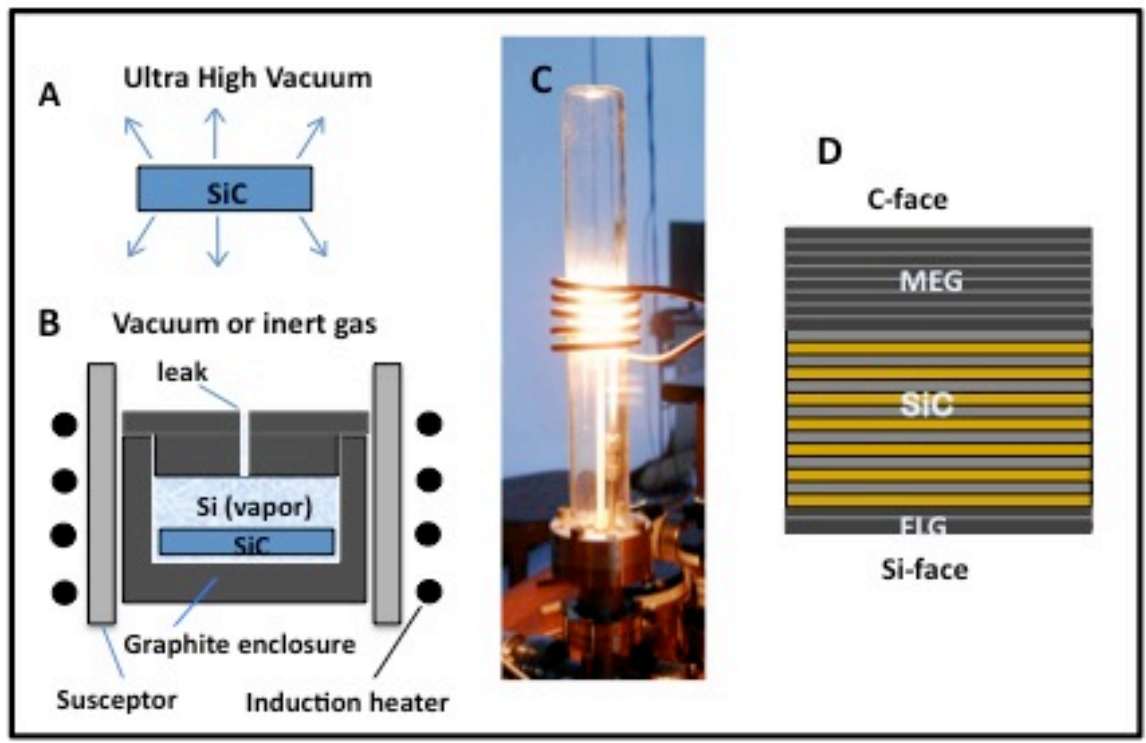

Figure 1

Figure 2

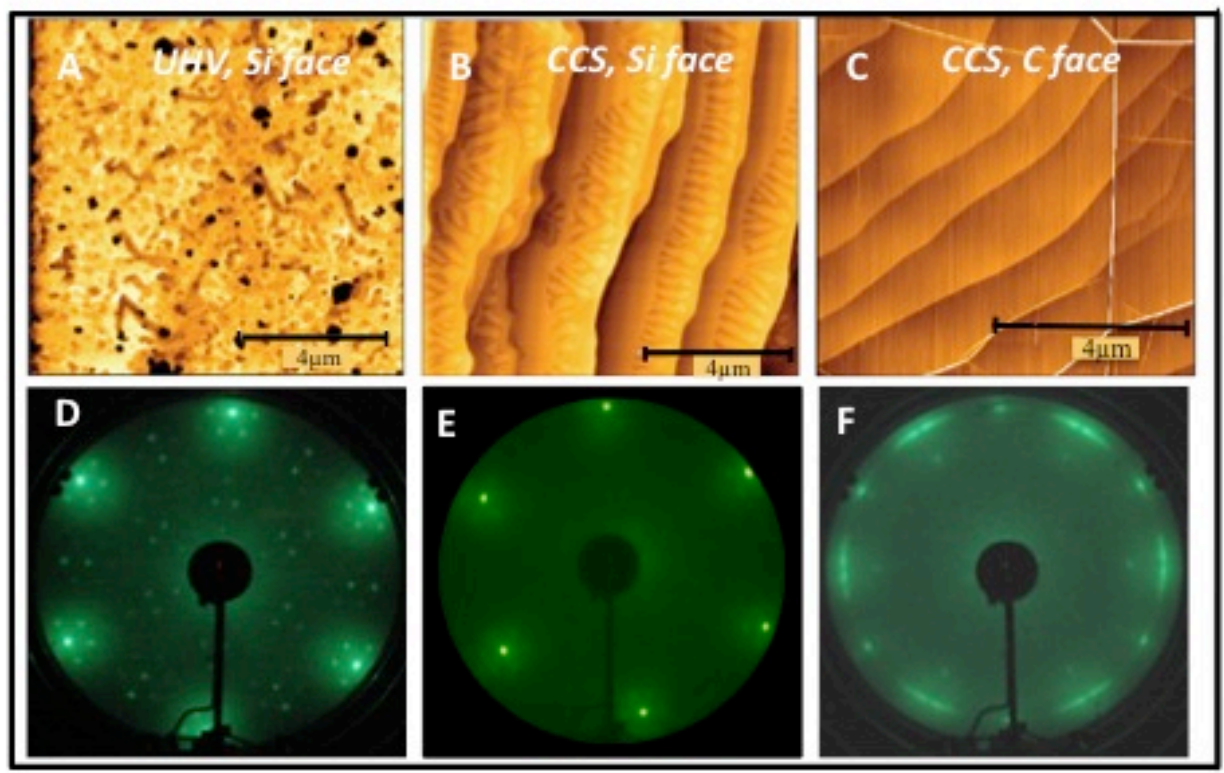

Figure 3

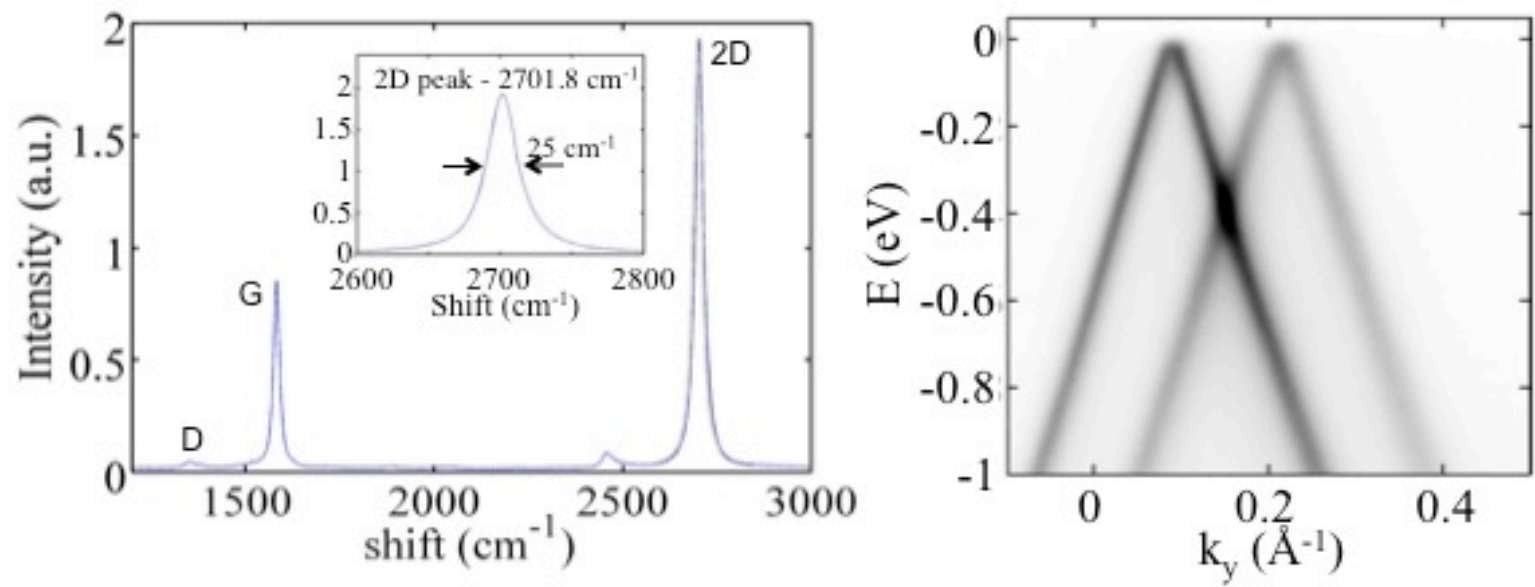

Figure 4

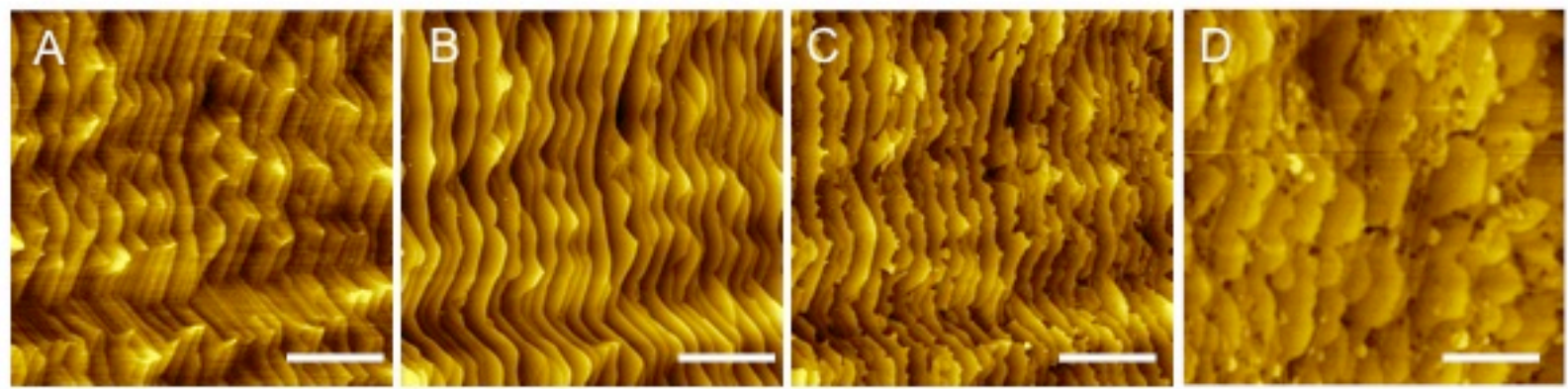

Figure 5

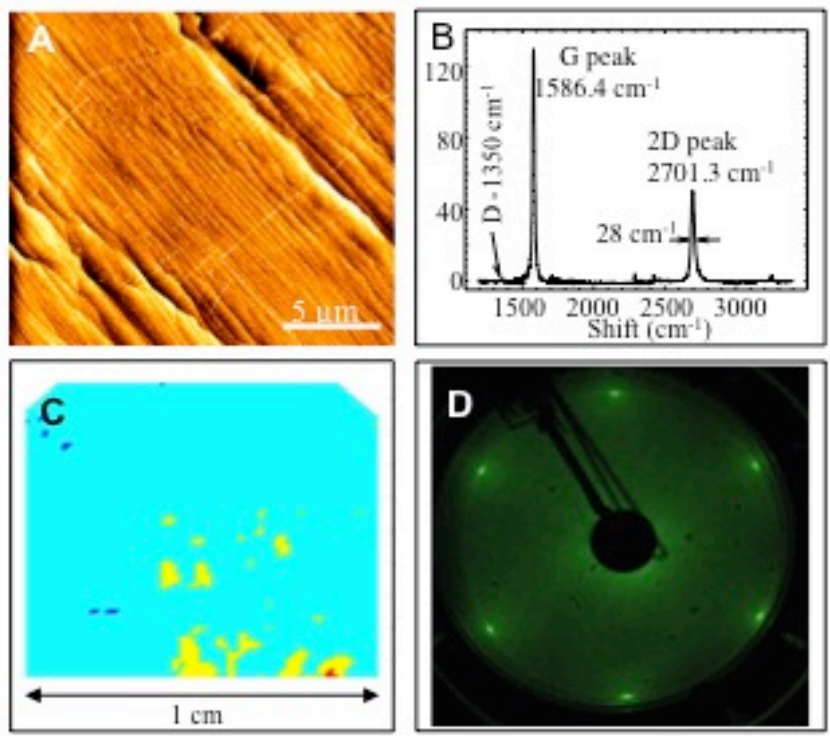

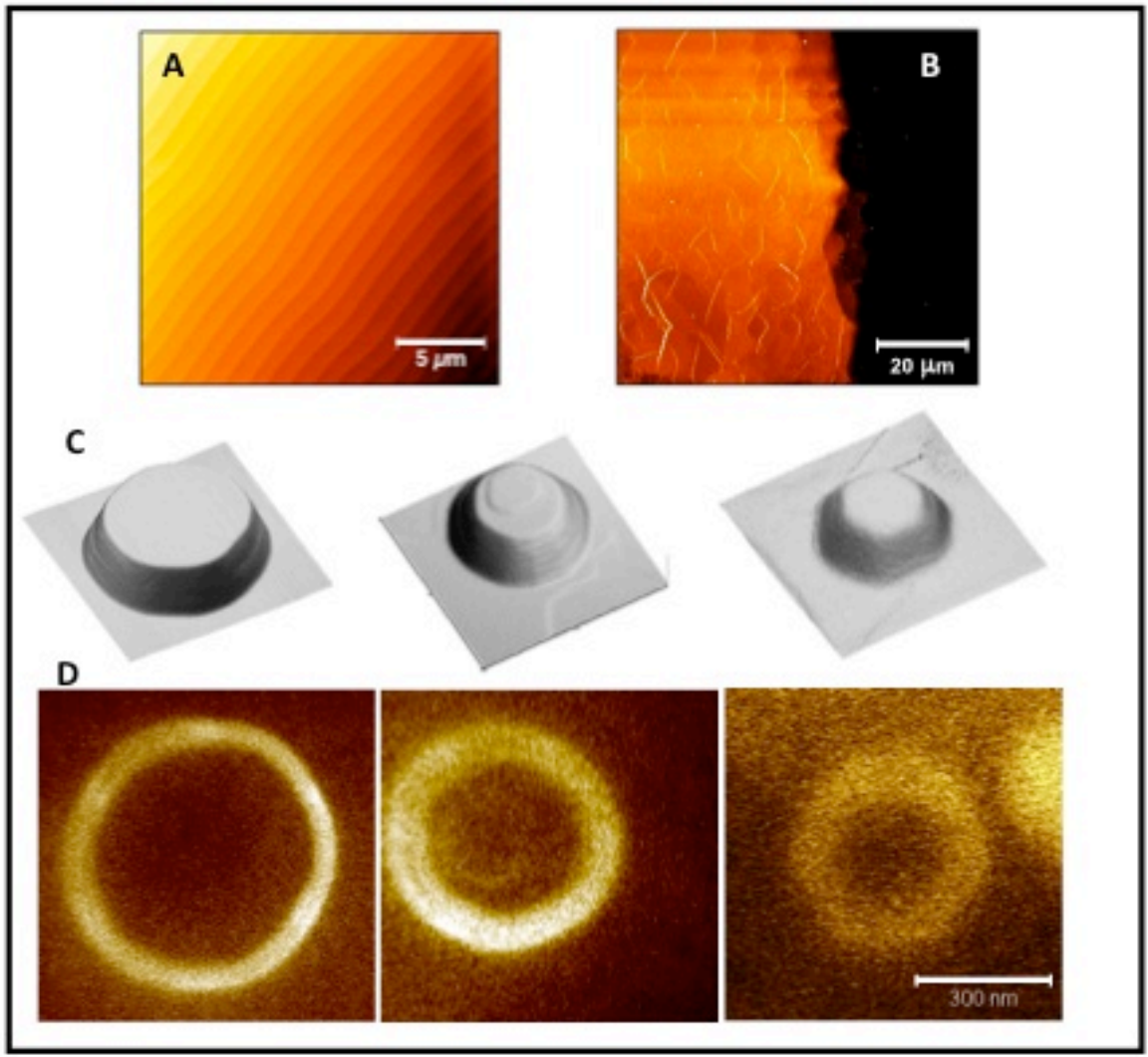

Figure 6

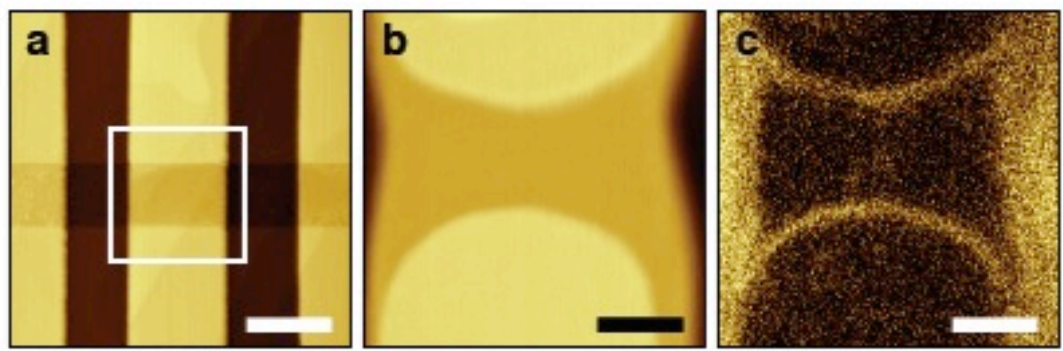
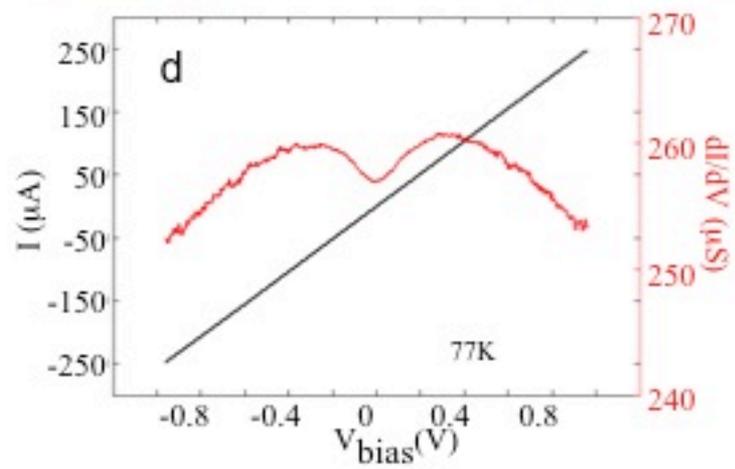

Figure 7